\documentclass[aps,prl,twocolumn,preprintnumbers,superscriptaddress,groupedaddress,nofootinbib]{revtex4}
\pdfoutput=1
\usepackage{graphicx}     %usepackage without driver option

\usepackage[bookmarksopen,colorlinks=true,linkcolor=steelblue,
citecolor=orangered,urlcolor=darkred,linktoc=all]{hyperref}

\usepackage{amsmath}
\usepackage{amsfonts}
\usepackage{amssymb}
\usepackage{bm}        % for math
\usepackage{slashed}   % for math

\usepackage{amsopn}
\mathchardef\mhyphen="2D

\usepackage{xcolor}
\definecolor{darkred}{rgb}{0.7, 0., 0.}
\definecolor{orangered}{rgb}{1,0.27,0.}
\definecolor{steelblue}{rgb}{0.275,0.51, 0.706}
\definecolor{forestgreen}{rgb}{0.13,0.55,0.13}

\newcommand{\abs}[1]{\left\lvert #1 \right\rvert}

\begin{document}
%%%%%%%%%%%%%%%%%%%%%%%%%%%%%%%%%%%%%%%%%%%%%%%%

%%%%%%%%%%%%%%%%%%%%%%%%%%%%%%%%%%%%%%%%%%%%%%%%
\preprint{DESY 21-118}

\title{\Large Improved indirect limits on muon EDM}

\author{Yohei Ema}
\email{yohei.ema@desy.de}
\affiliation{Deutsches Elektronen-Synchrotron DESY, Notkestr. 85, 22607 Hamburg, Germany}
\author{Ting Gao}
\email{gao00212@umn.edu}
\affiliation{School of Physics and Astronomy, University of Minnesota, Minneapolis, MN 55455, USA}
\author{Maxim Pospelov}
\email{pospelov@umn.edu}
\affiliation{School of Physics and Astronomy, University of Minnesota, Minneapolis, MN 55455, USA}
\affiliation{William I. Fine Theoretical Physics Institute, School of Physics and Astronomy,
University of Minnesota, Minneapolis, MN 55455, USA}

\date{\today}

\begin{abstract}
Given current discrepancy in muon $g-2$ and future dedicated efforts to measure muon electric dipole moment (EDM) $d_\mu$, we assess the indirect constraints imposed on $d_\mu$ by the EDM measurements performed with heavy atoms and molecules. We notice that the dominant muon EDM effect arises via the muon-loop induced ``light-by-light" $CP$-odd amplitude $\propto{\bf B}{\bf E}^3$, and in the vicinity of a large nucleus the corresponding parameter of expansion can be significant, $eE_{\rm nucl}/m_\mu^2 \sim 0.04$. We compute the $d_\mu$-induced Schiff moment of the $^{199}$Hg nucleus, and the linear combination of $d_e$ and semileptonic $C_S$ operator (dominant in this case) that determine the $CP$-odd effects in ThO molecule. The results, $d_\mu(^{199}{\rm Hg}) < 6\times 10^{-20}e$cm and $d_\mu({\rm ThO}) < 2\times 10^{-20}e$cm, constitute approximately three- and nine-fold improvements over the limits on $d_\mu$ extracted from the BNL muon beam experiment. 
\end{abstract}

%\keywords{}

\maketitle
%%%%%%%%%%%%%%%%%%%%%%%%%%%%%%%%%%%%%%%%%%%%%%%%

%%%%%%%%%%%%%%%
\textbf{Introduction.}\,---\,
The searches for EDMs of elementary particles progressed a long way since the first indirect limit on neutron EDM found by Purcell and Ramsey seventy years ago~\cite{Purcell:1950zz}. 
Current precision improved by nearly ten orders of magnitude since~\cite{Purcell:1950zz} and nil results of the most precise measurements~\cite{Graner:2016ses,Cairncross:2017fip,Andreev:2018ayy,nEDM:2020crw} have served a death warrant to many models that seek to break $CP$ symmetry at the weak scale in a substantial way (see {\em e.g}~\cite{Khriplovich:1997ga,Ginges:2003qt,Pospelov:2005pr,Engel:2013lsa}). 

EDMs of neutron and heavy atoms can also serve to constrain EDMs of heavier particles that do not appear inside these light objects ``on-shell"~\cite{Marciano:1986eh}. While for the EDMs (and color EDMs) of heavy quarks the gluon mediation (and for heaviest objects such as $t$-quark, Higgs mediation) diagrams play a crucial role~\cite{Weinberg:1989dx,Barr:1990vd}, the EDMs of muons and $\tau$-leptons require three-loop $\alpha_{\rm EM}^3$ suppressed amplitudes to generate the electron EDM $d_e$ via radiative corrections \cite{Grozin:2008nw}. In this work, we re-evaluate the muon EDM ($d_\mu$) induced $CP$-odd observables and find the enhanced sensitivity to $d_\mu$ in experiments that measure EDMs of heavy atoms/molecules. 

Latest interest to muons is fueled by the on-going discrepancy between theoretical predictions and experimental measurement of the muon anomalous magnetic moment~\cite{Davier:2017zfy,Colangelo:2018mtw,Hoferichter:2019mqg,Davier:2019can,Keshavarzi:2019abf,Aoyama:2020ynm,Muong-2:2021ojo}. It brings into focus a question of other observables that involve muons, and one such important quantity is the muon EDM, $d_\mu$ (see \emph{e.g.}~\cite{Crivellin:2018qmi} on extended discussion on this point). At the moment, the auxiliary EDM measurement at the Brookhaven $g-2$ experiment sets the tightest bound on muon EDM~\cite{Muong-2:2008ebm},
\begin{equation}
    |d_\mu| < 1.8\times 10^{-19}\,e{\rm cm},
    \label{BNL}
\end{equation}
but there are proposals on significantly improving this bound with dedicated muon beam experiments~\cite{Semertzidis:1999kv,Iinuma:2016zfu,Abe:2019thb,Adelmann:2021udj}. Given these upcoming efforts it is important to re-evaluate {\em indirect} bounds on muon EDM, especially given significant progress in precision of atomic/molecular EDM experiments in recent years. 

In this {\em Letter} we evaluate indirect limits on $d_\mu$ finding superior bounds to~(\ref{BNL}) from Hg and ThO EDM experiments~\cite{Graner:2016ses,Andreev:2018ayy}. Our results draw heavily on the fact that the closed muon loop with $d_\mu$ insertion is placed in a very strong electric field of a large nucleus ({\em e.g.} Hg or Th).  The resulting interaction, encapsulated by ${\bf E}^3{\bf B}$ effective operator, is capable of generating Schiff moment~\cite{Schiff:1963zz}, $CP$-odd electron-nucleus interaction~\cite{Khriplovich:1997ga}, and magnetic quadrupole moment. Below, we elaborate on details of our findings.

\textbf{Muon EDM and $E^3B$ interaction.}\,---\,
The input into our calculations is the muon EDM operator,
\begin{equation}
\label{dmu}
    {\cal L}_{CP\mhyphen \mathrm{odd}} = -\frac{i}{2} F^{\alpha\beta}\times  \overline{\mu} \sigma_{\alpha\beta}\gamma_5 \mu \times d_\mu,
\end{equation}
and for the purpose of this paper we assume that the Wilson coefficient $d_\mu$ is the only source of $CP$-violation. 

\begin{figure}[t]
  \includegraphics[scale=1]{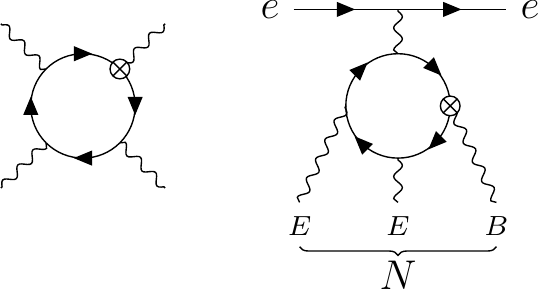}
  \caption{A representative light-by-light scattering diagram with $d_\mu$ insertion (indicated by the crossed dot) giving rise to $E^3B$ interaction. When $E^2B$ is sourced by the nucleus, as shown on the right, $d_N$ and $S_N$ are generated.} 
   \label{CP-EH}
\end{figure}

At one loop order,  muons induce $CP$-odd nonlinear electromagnetic interactions, much the same as the well-studied ``light-by-light" diagrams in the $CP$-even channel. In Fig.~\ref{CP-EH} we show an example of such diagram. We notice that photon momenta entering the muon loop are small compared to the muon mass $m_\mu$. Indeed, in a large nucleus, $q_\gamma^{\rm max}\sim R_N^{-1}\sim 30$\,MeV, one can truncate the series to the lowest dimension operator, and assume electric ${\bf E}$ and magnetic ${\bf B}$ fields to be uniform. Working in the lowest order in $d_\mu$, we directly compute the corresponding electromagnetic operators, similar to the dimension eight  term in the Euler-Heisenberg Lagrangian:
\begin{align}
\label{F3Fdual}
    {\cal L}_{\rm } &= - e^4(\tilde F_{\alpha\beta} F^{\alpha\beta})(F_{\gamma\delta}  F^{\gamma \delta}) \times \frac{d_\mu/e}{96 \pi^2m_\mu^3}\nonumber \\
    &= -\frac{d_\mu/e}{12 \pi^2m_\mu^3}e^4({\bf E\cdot B})({\bf E\cdot E}-{\bf B\cdot B}),
\end{align}
where $\tilde F_{\alpha\beta} = \frac12 \epsilon_{\alpha\beta\mu\nu}F^{\mu\nu}$, and we define the gauge coupling $e$ to be positive.
One can notice interesting differences with $CP$-even case: dimension four $(\tilde F_{\alpha\beta} F^{\alpha\beta})$ operator can be dropped, and there is only one dimension eight operator $(FF)(F\tilde F)$, while $CP$-even case has two, $(FF)(FF)$ and $(F\tilde F)(F\tilde F)$. 
The effective $CP$-odd photon interactions were discussed recently in \cite{Gorghetto:2021luj}.
In principle, all terms in the expansion can be computed analytically. Neglecting $O(B^3)$ interaction
that is subdominant due to no $Z$-enhancement leaves only $E^3B$ effective operator that we write in a more generic form that can be applied to other sources of $CP$-violation as well: 
\begin{equation}
\label{E3B}
H_\mathrm{eff}  = C_{E^3B}\times \int d^3x \,e^4({\bf E\cdot E})({\bf E\cdot B}),
\end{equation}
with $C_{E^3B}= (12 \pi^2m_\mu^3)^{-1}d_\mu/e$ in our model~(\ref{dmu}).

It is important to note that 
the $E^3B$ effective interaction does not always capture all relevant physics. For example, the muon-loop-mediated electron EDM that arizes at three loop order involves computation with loop momenta that can be comparable or even larger than $m_\mu$. 
In that case, the entire $CP$-odd four-photon amplitude is needed~\cite{Grozin:2008nw}. 
In what follows we evaluate the physical consequences of the $E^3B$ interaction.

\textbf{Muon EDM and nuclear $CP$-odd observables}\,---\, Nuclear spin dependent EDMs (sometimes called diamagnetic EDMs) provide stringent tests of $CP$-violation via probing nuclear $T,P$-odd moments. At this step we address the mechanisms that convert $CP$-even static nuclear moments to the $CP$-odd ones,
\begin{equation}
    \mu_N, Q_N ~\xrightarrow{E^3B} ~ d_N, S_N, M_N,
\end{equation}
where subscript $N$ stands for ``nuclear", and $\mu,\,Q,\,d,\,S,\,M$ are magnetic, electric quadrupole, electric dipole, Schiff and magnetic quadrupole moments. (Inside a neutral atom, $d_N$ is not observable by itself, but in the linear combination that parametrizes the difference between EDM and charge distribution, the Schiff moment~\cite{Schiff:1963zz}.)

Consider a spin-$\frac12$ nucleus, as in the most sensitive diamagnetic EDM experiment with $^{199}$Hg
\cite{Graner:2016ses}. Then $M_N$ is absent by definition, but $d_N$ and $S_N$ can be induced as shown in Fig.~\ref{CP-EH}. To calculate them we notice that the magnetic field of the $I=1/2$ nucleus can be presented in the following form:
\begin{equation}
\label{B}
    eB_{i}({\bf r})= b_1(r)n_{Ii} + b_2(r)(3n_in_j-\delta_{ij}) n_{Ij},
\end{equation}
where we introduced the unit vector in the direction of the nuclear spin, ${\bf n}_{I} = {\bf I}/I$, ${\bf n}= {\bf r}/r$ and some scalar invariant functions $b_{1(2)}(r)$. Notice that in the limit of a very small nuclear radius, $R_N \to 0$, the corresponding asymptotics of these functions are
\begin{equation}
\label{bc}
b_1(r) \to \frac{2 e\mu_N}{3}\delta({\bf r});~~
b_2(r) \to \frac{e\mu_N}{4\pi r^3}.
\end{equation}
where $\mu_N$ is the nuclear magnetic dipole moment value. 
The nuclear electric field, to good accuracy, can be described by the radial ansatz, 
\begin{equation}
\label{E}
    e{\bf E} = \frac{{\bf n}}{r^2}\times Z\alpha f(r),
\end{equation}
where $Z$ is the atomic number, $\alpha$ is the fine structure constant and 
$f(r)$ is the fraction of nuclear charge within the radius $r$. For the uniform sphere charge distribution $f(r) = r^3/R_N^3$ for $r<R_N$ and $f(r)=1$ for $r>R_N$. Substituting~(\ref{E}) and~(\ref{B}) into~(\ref{E3B}) and performing angular integration, we obtain intermediate expressions for $d_N$ and $S_N$:
\begin{alignat}{2}
\label{dN}
\frac{d_N}{eC_{E^3B}} &= 4\pi(Z\alpha)^2 \int \frac{dr }{r^2}f^2&&\left(\frac53b_1 +\frac43b_2\right),\\
\frac{S_N}{eC_{E^3B}} &= \frac{2\pi(Z\alpha)^2}{15} \int dr f^2&&\left[b_1\left(11-\frac{25}{3}\frac{r_c^2}{r^2} \right)\right. \nonumber
\\ &&&\left. +b_2\left(16-\frac{20}{3}\frac{r_c^2}{r^2} \right) \right].
\label{SN}
\end{alignat}
In these expressions, $r_c^2$ is the nuclear charge radius. We follow the standard definition of the Schiff moment that in non-relativistic limit and point-like nucleus leads to the effective nuclear-spin-dependent $T,P$-odd Hamiltonian for electrons 
\begin{equation}
    H_{T,P\mhyphen\mathrm{odd}}  = - (S_N/e) \times 4\pi \alpha ({\bf n}_I\cdot \bm{\nabla}_e)\delta({\bf r}_e).
\end{equation}
Nuclear dependence in~(\ref{dN}) and~(\ref{SN}) is encapsulated in~$f$ and~$b_i$. Electric field, {\em i.e.} $f$, is determined by the collective properties of the nucleus and has little to no dependence on the details of the nucleon's wave function inside a large nucleus. In contrast, the scalar functions~$b_i$ that describe magnetization are determined by mostly ``outside" valence nucleons and carry more detail about nuclear structure. For any realistic choice of $f$ and $b_i$, however, it is easy to see that radial integrals will be saturated by distances $r\sim R_N$.

Specializing our calculations to the $^{199}$Hg nucleus, we adopt a simple shell model description of it with a valence neutron in $n_r=2,\,l=1,\,j=1/2$ state carrying all angular momentum dependence, and ignore configuration mixing. Its wave function can be conveniently written as
\begin{equation}
    \psi({\bf r}_n) = R_{2p}(r_n)\frac{(\bm{\sigma}_n\cdot {\bf n}_n)}{\sqrt{4\pi}}\chi,
    \label{eq:valence_neutron_wf}
\end{equation}
where ${\bf r}_n = {\bf n}_nr_n$ and $\chi$ are neutron's coordinate and two component spinor, and $R_{2p}$ is the radial wave function normalized as $\int R^2r^2dr=1$.
Nuclear spin in this case coincides with $j$, and 
${\bf n}_I = \chi^\dagger \bm{\sigma}_n \chi $.
The magnetic moment of the nucleus has a simple connection to the magnetic moment of the neutron,
$e \mu_N=(-1/3)e\mu_n= (-1/3) \times (-1.91) \times 4\pi\alpha/(2m_p)$. The magnetization functions $b_i$ defined earlier in~(\ref{B}) can be directly related to radial $R_{2p}$ functions, and explicit calculations give
\begin{align}
b_1(r) &= \frac{-1.91\alpha}{2m_p}\times \frac{2}{3}  \left(2\int_r^\infty \frac{dr_n}{r_n}R_{2p}^2(r_n)-R_{2p}^2(r) \right), \nonumber\\
b_2(r) &= \frac{-1.91\alpha}{2m_p}\times\frac{1}{3}\left(R_{2p}^2(r) - \frac{1}{r^3} \int_0^r dr_n r_n^2R_{2p}^2(r_n)  \right).
\nonumber
\end{align}
One can easily check that the corresponding boundary conditions~(\ref{bc}) are satisfied. To learn about the parametric dependence of our answers we first explore the simplified case when not only the charge distribution but also $R(r)$ is taken to be constant inside the nuclear radius and zero outside, $R_{2p}^2(r)= 3R_N^{-3}\theta(R_N-r)$~\cite{Ginges:2003qt}. In this approximation we get
\begin{equation}
    \frac{d_N}{eC_{E^3B}} = \frac{1.91\times 2\pi Z^2\alpha^3}{3m_pR_N^4};~ 
    \frac{S_N}{eC_{E^3B}} = \frac{1.91\times 39\pi Z^2\alpha^3}{245m_pR_N^2},
    \label{naive}
\end{equation}
and consequently $S_N$ scales as $Z^{4/3}$ since $R_N \propto Z^{1/3}$. 
In order to get a more realistic answer, we solve for $R_{2p}$ numerically using the Woods-Saxon potential with parameters outlined in Ref.~\cite{Dmitriev:2003sc}. We check that our results reproduce $S_N(d_n)$~\cite{Dmitriev:2003sc,Ginges:2003qt} with reasonable $\propto 30\%$ accuracy. Performing two numerical integrals over $r_n$ and $r$, and substituting explicit expression for $C_{E^3B}$, we obtain the following numerical result, 
\begin{equation}
\label{numerical}
    S_{^{199}{\rm Hg}}/e \simeq (d_\mu/e)\times 4.9\times 10^{-7}\, {\rm fm}^2, 
\end{equation}
that lands itself very close (withing 20\%) from the naive estimate~(\ref{naive}). 
Given the experimental constraint of 
$|S_{^{199}{\rm Hg}}| < 3.1\times 10^{-13} \,e\, {\rm fm}^3$~\cite{Graner:2016ses}, we arrive at the following final result
\begin{equation}
\label{Sbound}
    |d_\mu| < 6.4\times 10^{-20}\,e\,{\rm cm},
\end{equation}
which is somewhat more stringent bound, by a factor of $\sim 2.5$ than~(\ref{BNL}). Result (\ref{numerical}) carries a 25-30\% uncertainty due to neglected contributions from the nuclear orbital mixing. 

Future developments may bring about new experiments that would search for EDMs involving nuclei with $I\geq 1$~\cite{Flambaum:2014jta}, opening the possibility of measuring magnetic quadrupole moments, and using nuclei with large deformations/large $Q_N$. 
We perform a simple estimate for the expected size of the magnetic quadrupole by taking the electric field created by $Q_N$ outside the nucleus, and cutting divergent integrals at $R_N$. This way, we arrive at the following estimate
\begin{equation}
    \frac{M_N}{eC_{E^3B}} \sim  
     \frac{48\pi Z^2\alpha^3}{5}    \frac{Q_N}{e}\int \frac{dr}{r^5}\simeq  \frac{Q_N}{e} \frac{12\pi Z^2\alpha^3}{5R_N^4}.
\end{equation}
Substituting expression~(\ref{E3B}), and normalizing electric quadrupole on large values observed in deformed nuclei, we get
\begin{equation}
 \frac{M_N}{e}    \sim 10^{-4}\,{\rm fm}\times \frac{Q_N}{e\, 300\,{\rm fm}^2}\times  (d_\mu/e).
\end{equation}
Taking typical matrix elements and extrapolating future sensitivity to the current one of the ThO experiment, one could probe $M_N/e\propto 10^{-11}\,{\rm fm}^2$ and consequently achieving $d_\mu/e\propto 10^{-20}\, e\,{\rm cm}$.

\textbf{Muon EDM and paramagnetic $CP$-odd observables.}\,---\, Finally we turn our attention to the electron-spin-dependent EDMs referred to as paramagnetic EDMs of atoms and molecules. These experiments probe the electron EDM operator (defined through Eq.~(\ref{dmu}) with $\mu\to e$) and semi-leptonic $CP$-odd operators among which the most important one  is $C_S$, 
\begin{equation}
    {\cal L}_{eN} = C_S\frac{G_F}{\sqrt{2}}(\bar e i \gamma_5 e) ( \bar pp + \bar nn).
    \label{eq:CS}
\end{equation}
For non-relativistic electrons and small $R_N$ limit, this term gives rise to $\propto ({\bm \sigma}_e\cdot \bm{\nabla}_e)\delta({\bf r}_e)$ effective interaction. 
The importance of $C_S$ for probing $CP$ violation in the Higgs sector, quark sector etc has been emphasized many times in the literature, see {\em e.g.}~\cite{Barr:1991yx,Lebedev:2002ne,Jung:2013hka,Flambaum:2019ejc}. Tremendous progress of the past decade with limits on $d_e$ and $C_S$ has been achieved by the ACME collaboration in experiment with the ThO paramagnetic molecule~\cite{Andreev:2018ayy}. Since the results are often reported in terms of $d_e$, it is convenient to introduce a linear combination of the two quantities 
limited in experiment and refer to them as ``equivalent $d_e" $\cite{Pospelov:2013sca}:\footnote{
	The sign convention of $C_S$ can be checked, \emph{e.g.}, with~\cite{Dzuba:2011}.
	We define $\gamma_5 = i\gamma^0 \gamma^1 \gamma^2 \gamma^3$
	that has the opposite sign as theirs.
}
\begin{equation}
\label{deeq}
    d_e^{\rm equiv} = d_e +C_S\times 1.5\times 10^{-20}\,e\,{\rm cm}.
\end{equation}
Current experimental limit stands as 
$|d_e^{\rm equiv}| < 1.1 \times 10^{-29}\,e\, {\rm cm}$~\cite{Andreev:2018ayy}.

\begin{figure}
%\hspace*{-0.1in}
%\vspace*{0.1cm}
  \includegraphics[scale=1]{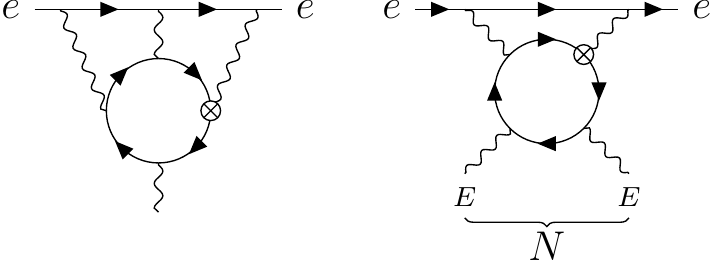}
  \caption{Three-loop contribution to $d_e$ and two-loop contribution to equivalent $C_S$ generated by $d_\mu$. } 
   \label{3loop}
\end{figure}

Muon EDM contributes both to $d_e$ and $C_S$ through loops. The bona fide three-loop $d_e(d_\mu)$ computation, Fig.~\ref{3loop}, was performed in~\cite{Grozin:2008nw},
\begin{equation}
\label{de}
d_e = d_\mu \left(\frac{\alpha}{\pi}\right)^3\frac{m_e}{m_\mu}\times 1.92 \simeq 1.1\times 10^{-10} d_\mu.
\end{equation}
If the direct bound~(\ref{BNL}) is saturated, $d_e$ will be larger than the experimental limit by about a factor of two, as already noted in Ref. \cite{Crivellin:2018qmi}. It turns out, however, that equivalent of $C_S$ generated by $E^3B$ interaction gives a larger contribution. 

A representative diagram contributing to the $T,P$-odd electron-nucleus interaction via $E^3B$ term is shown in Fig.~\ref{3loop}. The two electric field lines can be sourced by a nucleon, or a nucleus, while the photon loop attached to electron line generates $m_e\bar e i \gamma_5 e$ interaction. There are two important considerations regarding this type of contribution: {\em i.} The photon loop is enhanced by $\log(\Lambda/m_e)$, and we calculate this loop to logarithmic accuracy, cutting it at $\Lambda=m_\mu$. (In practice, this cutoff will be supplied by the non-local nature of the muon loop in Fig.\,\ref{CP-EH}.) {\em ii.} In a large nucleus ${\bf E}^2$ is coherently enhanced and dominates over effects proportional to electromagnetic contribution of individual nucleons $\propto Z\langle p| {\bf E}^2 |p\rangle$. Being concentrated  inside and near the nucleus, ${\bf E}^2$ can be considered {\em equivalent} to the delta-functional contribution:
\begin{equation}
\label{delta_eq}
    e^2 ({\bf E}^2)_{\rm nucl} \to \delta({\bf r})\times \frac{4\pi (Z\alpha)^2}{R_N}\times 
    \int_0^\infty\frac{ f^2(R_N x)}{x^2} dx,
\end{equation}
where $x=r/R_N$.
For a constant density charge distribution, the integral in~(\ref{delta_eq}) is 6/5, and we adopt this number. Putting the results of the loop calculation together with~(\ref{delta_eq}), and using the explicit form for $C_{E^3B}$ we arrive at the following prediction for the {\em equivalent} $C_S$ value: 
\begin{equation}
    \label{csfinal}
    \frac{G_F}{\sqrt{2}} C_S^{\rm equiv}= 
    \kappa \frac{4 Z^2\alpha^4}{\pi A} \times \frac{m_e(d_\mu/e)}{m_\mu^3R_N}\times 
    \log\left(\frac{m_\mu}{m_e}\right) 
    .
\end{equation}
As one can see, $C_S^{\rm equiv}$ scales as $Z^2A^{-1}R_N^{-1}\propto Z^{2/3}$, %since $R_N \propto Z^{1/3}$, 
which is the sign of coherent enhancement. $A$ is the number of nucleons, and $A= 232$ for Th.
In this expression, $\kappa$ is a fudge factor to account for the change of the electronic matrix elements stemming from the fact that nuclear ${\bf E}^2$ extends beyond the nuclear boundary, while true nucleonic $C_S$ effect is proportional to nuclear density and vanishes outside. Solving the Dirac equation near the nucleus for the outside $s_{1/2}$ and $p_{1/2}$ electron wave functions and finding a ratio of the matrix elements for these two distributions result in $\kappa \simeq 0.66$. We then arrive to the numerical result 
\begin{equation}
\label{Csnumbers}
    C_S^{\rm equiv} = 3.1\times 10^{-10} \left(\frac{d_\mu}{10^{-20}\,e\,{\rm cm}}\right).
\end{equation}
Combining~(\ref{Csnumbers}) with~(\ref{de})
into~(\ref{deeq}), we arrive at our main result 
\begin{equation}
\label{Csbound}
 d_e^{\rm equiv} \simeq 5.8\times 10^{-10} \, d_\mu~\Longrightarrow ~  |d_\mu| < 1.9\times 10^{-20}\,e\,{\rm cm}. 
\end{equation}
We observe that $d_e$ and $ C_S^{\rm equiv}$ interfere constructively, and $C_S$ contribution is larger by a factor of $\simeq 4$. We believe (\ref{Csnumbers})to be accurate within $\sim 15-20\%$ with uncertainties associated with modelling of ${\bf E}(r)$ and logarithmic approximation for the photon loop integral.

\textbf{Outlook}\,---\, We have evaluated the electromagnetic transmission mechanisms of muon EDM to the observable EDMs that do not involve on-shell muons. We have found that muon-loop-induced $E^3B$ effective interaction plays an important role and leads to novel indirect bounds, Eqs.~(\ref{Sbound}) and~(\ref{Csbound}) that are already stronger than the direct bound~(\ref{BNL}). Result~(\ref{Csbound}) provides a new benchmark that future dedicated muon EDM experiments would have to overtake. We also notice that since both $^{199}$Hg and ThO EDM results give an  improvement, it is highly unlikely that a fine-tuned choice of $d_e$ and hadronic $CP$-violation would lead to the relaxation of indirect bounds on $d_\mu$. 

In this paper, we do not discuss the short-distance physics that may lead to the enhanced $d_\mu$. We note that while in some models $d_\mu$ is predicted at the same level as $d_e$, it is also feasible that $d_\mu/d_e $ scales as $(m_\mu/m_e)^3$ and possibly even larger. (Given the on-going $g-2$ discrepancy in the muon sector, it is clear that $d_\mu$ deserves a separate treatment.) Still, it is instructive to equate $d_\mu$ to some simple scaling formula that involves an ultraviolet scale $\Lambda_\mu$, and we choose $d_\mu = m_\mu/\Lambda_\mu^2$ scaling. Then our results translate to 
\begin{equation}
    \Lambda_\mu > 300\,{\rm GeV},
\end{equation}
which underscores that the (weak scale)$^{-1}$ distances start being probed. Depending on underlying model, there can be some scale dependence of the muon EDM form factor $d_\mu({Q^2})$ (see {\em e.g.} \cite{Grozin:2008nw}). This, however, does not obscure comparison of direct ($Q^2\simeq 0$) and indirect ($Q^2\simeq m_\mu^2$) limits derived in our paper as long as $d_\mu$ operator is generated at distances $\Lambda^{-1}\ll m_{\mu}^{-1}$. 

We also update the limit on the $\tau$-lepton EDM $d_\tau$ derived in \cite{Grozin:2008nw}. Our analysis is directly applicable to $d_\tau$
after replacing $m_\mu$ by the $\tau$-lepton mass $m_\tau$. 
In this case, the electron EDM plays the dominant role since $d_e \propto m_\tau^{-1}$ while $S_N, C_S \propto m_\tau^{-3}$
up to logarithm.
For the ThO molecule, we obtain
\begin{equation}
 d_e^{\rm equiv} \simeq 7.0\times10^{-12} \, d_\tau~\Longrightarrow ~  |d_\tau| < 1.6\times 10^{-18}\,e\,{\rm cm}.
 \label{eq:tauEDM}
\end{equation}
This surpasses the constraint from the Belle experiment~\cite{Belle:2002nla}.
The constraint from ${}^{199}$Hg is weaker by a factor of $\sim 2\times 10^2$ than~\eqref{eq:tauEDM}.

Finally, while the focus of our paper was on $d_\mu$, one could also derive limits on $C_{E^3B}$ applicable to other models. We get constraints on  $C_{E^3B}$ at the level of $10^{-41}\,{\rm eV}^{-4}$ and better, which would be challenging to match with photon-based experiments \cite{Gorghetto:2021luj}.

%%%%%%%%%%%%%%%

%%%%%%%%%%%%%%%%%%%%%%%%%%%%%%%%%%%%%%%%%%%%%%%%
\vspace{3.5mm}
\begin{acknowledgments}
\textbf{Acknowledgments}\,---\,
This work was partly funded by the Deutsche Forschungsgemeinschaft under Germany’s Excellence Strategy - EXC 2121 “Quantum Universe” - 390833306.
M.P. is supported in part by U.S. Department of Energy Grant No.
desc0011842. The Feynman diagrams in this paper are drawn with \texttt{TikZ-Feynman}~\cite{Ellis:2016jkw}.
\end{acknowledgments}
%%%%%%%%%%%%%%%%%%%%%%%%%%%%%%%%%%%%%%%%%%%%%%%%

%%%%%%%%%%%%%%%%%%%%%%%%%%%%%%%%%%%%%%%%%%%%%%%%
\bibliographystyle{utphys}
\bibliography{muonEDM}
%%%%%%%%%%%%%%%%%%%%%%%%%%%%%%%%%%%%%%%%%%%%%%%%

%%%%%%%%%%%%%%%%%%%%%%%%%%%%%%%%%%%%%%%%%%%%%%%%
\clearpage
\appendix
\onecolumngrid
%%%%%%%%%%%%%%%%%%%%%%%%%%%%%%%%%%%%%%%%%%%%%%%%

\renewcommand{\thesection}{S\arabic{section}}
\renewcommand{\theequation}{S\arabic{equation}}
\renewcommand{\thefigure}{S\arabic{figure}}
\renewcommand{\thetable}{S\arabic{table}}
\renewcommand{\thepage}{S\arabic{page}}
\setcounter{equation}{0}
\setcounter{figure}{0}
\setcounter{table}{0}
\setcounter{page}{1}

%%%%%%%%%%%%%%%%%%%%%%%%%%%%%%%%%%%%%%%%%%%%%%%%
\begin{center}
\textbf{\Large Supplemental Material
}
\end{center}
%%%%%%%%%%%%%%%%%%%%%%%%%%%%%%%%%%%%%%%%%%%%%%%%

In this Supplemental Material, we provide technical details on
the evaluation of the Schiff moment and the semi-leptonic $CP$-odd operator, as well as estimate theoretical errors in evaluating these quantitites.

%%%%%%%%%%%%%%%%%%%%%%%%%%%%%%%%%%%%%%%%%%%%%%%%
\section{Schiff moment}
%%%%%%%%%%%%%%%%%%%%%%%%%%%%%%%%%%%%%%%%%%%%%%%%
Here we start from the $CP$-odd photon operator~\eqref{E3B} and derive the Schiff moment~\eqref{SN}.
We focus on the part linear in the electric field induced by the electron as shown in Fig.~\ref{CP-EH}.
The $E^3B$ operator is then evaluated as
\begin{align}
	{H}_\mathrm{eff} &=
	-C_{E^3B} \times
	\int d^3 x \left(\bm{\nabla}\frac{\alpha}{\abs{\mathbf{x} - \mathbf{r}_e}}\right)\cdot
	\left(2e\mathbf{E} \left(e\mathbf{E} \cdot e\mathbf{B}\right) 
	+ e\mathbf{B}\left(e\mathbf{E} \cdot e\mathbf{E}\right)\right),
	\label{eq:Heff_before}
\end{align}
where $\mathbf{E}$ and $\mathbf{B}$ in this expression are understood to be the nuclear electromagnetic field,
and $\mathbf{r}_e$ is the position vector of the electron. 
With Eqs.~\eqref{B} and~\eqref{E}, we obtain
\begin{align}
	{H}_\mathrm{eff} &=
	\int d^3 x \left(\bm{\nabla}_e\frac{\alpha}{\abs{\mathbf{x} - \mathbf{r}_e}}\right)\cdot \mathbf{P}_d,
\end{align}
where the nuclear EDM distribution is given by
\begin{align}
	\mathbf{P}_d &=
	C_{E^3B}\frac{Z^2 \alpha^2 f^2}{r^4}
	\left[\frac{5b_1+4b_2}{3}\mathbf{n}_I + \frac{2b_1 + 7b_2}{3}
	\left(3(\mathbf{n}\cdot\mathbf{n}_I)\mathbf{n} - \mathbf{n}_I\right)
	\right].
\end{align}
We thus obtain the nuclear EDM as
\begin{align}
	\frac{\mathbf{d}_N}{e} &= \int d^3 x\,\mathbf{P}_d
	= \mathbf{n}_I \times C_{E^3B}\int d^3x
	\frac{Z^2 \alpha^2 f^2}{r^4}
	\left(\frac{5b_1+4b_2}{3}\right),
\end{align}
reproducing Eq.~\eqref{dN}.
Due to the screening effect, the atomic EDM is induced not solely by the nuclear EDM distribution but
by the interaction of the form
\begin{align}
	{H}_\mathrm{eff} &=
	\int d^3 x \left(\bm{\nabla}_e\frac{\alpha}{\abs{\mathbf{x} - \mathbf{r}_e}}\right)\cdot 
	\left(\mathbf{P}_d - \rho_q \frac{\mathbf{d}_N}{e}\right),
\end{align}
where $\rho_q$ is the nuclear charge distribution normalized as $\int d^3x \rho_q = 1$.
Since the atomic scale is much larger than the nuclear scale,
we may expand the electric field induced by the electron as
\begin{align}
	\bm{\nabla}_{e}\frac{1}{\abs{\mathbf{x}-\mathbf{r}_e}}
	= \bm{\nabla}_{e}\left[\frac{1}{r_e} - \mathbf{x}\cdot \bm{\nabla}_e \frac{1}{r_e} 
	+ \frac{1}{2}\left(\mathbf{x}\cdot\bm{\nabla}_e\right)^2\frac{1}{r_e} + \cdots \right].
\end{align}
The first two terms do not contribute and we obtain to the leading order 
\begin{align}
	H_\mathrm{eff} &= \frac{\alpha}{2}\left(\nabla_{i} \nabla_{j} \nabla_{k} \frac{1}{r_e}\right)
	\int d^3x \left[(P_{d})_i - \rho_q \frac{d_{Ni}}{e} \right] x_j x_k,
\end{align}
where we omit the subscript $e$ from $\nabla$ for notational ease
but the derivatives still act on $r_e$ as the bracket indicates.
After the angular integration, we obtain
\begin{align}
	H_\mathrm{eff}
	&= -\frac{S_N}{e} \times 4\pi \alpha \left(\mathbf{n}_I \cdot \bm{\nabla}_e\right) \delta(\mathbf{r}_e),
\end{align}
where the Schiff moment is given by Eq.~\eqref{SN}.

Up until this point, the treatment was completely general, and used only the symmetry considerations applied to ${\bf E}$ and ${\bf B}$. To move further and evaluate the Schiff moment, we adopt the model where ${\bf E}$ is created collectively by all protons inside the nucleus, while ${\bf B}$ is generated by a valence nucleon in a shell model of the nucleus. Evaluations of the magnetic moment of the $^{199}$Hg show that the latter approximation holds to $\sim20$\% accuracy. 
In our evaluation, we simply take $f(r) = r^3/R_N^3$ for $r<R_N$ and $f(r)=1$ for $r>R_N$
for the nuclear electric field. We have checked that
the result is affected only within $10\%$ if we instead use the Woods-Saxon type charge distribution.
The nuclear magnetic field induced by
the magnetic moment of the valence neutron is given by~\cite{Landau:1991wop} (notice the different normalization of $e$)
\begin{align}
	e\mathbf{B}(\mathbf{x}) &= \frac{e\mu_n}{4\pi} \int d^3 x_n 
	\left[\bm{\nabla}_n \times \left(\psi_n^\dagger(\mathbf{x}_n) \bm{\sigma}_n\psi_n(\mathbf{x}_n)\right)\right]
	\times \bm{\nabla}_n\frac{1}{\abs{\mathbf{x}_n-\mathbf{x}}},
\end{align}
where $\mathbf{x}_n$ is the position vector of the valence neutron and $\bm{\sigma}_n$ is the Pauli matrix.
The wave function of the valence neutron $\psi_n$ is normalized as
\begin{align}
	\int d^3x_n \abs{\psi_n}^2 = 1.
\end{align}
After integration by parts we obtain
\begin{align}
	e\mathbf{B}(\mathbf{x}) &= \frac{2 e\mu_n}{3}\psi_n^\dagger(\mathbf{x}) \bm{\sigma}_n \psi_n(\mathbf{x})
	+ \frac{e\mu_n}{4\pi}\left[\bm{\nabla}\left(\bm{\nabla} \cdot \right) - \frac{\bm{\nabla}^2}{3}\right]
	\int d^3 x_n \frac{\psi_n^\dagger(\mathbf{x}_n) \bm{\sigma}_n \psi_n(\mathbf{x}_n)}{\abs{\mathbf{x}_n - \mathbf{x}}}.
	\label{eq:Bfield}
\end{align}
With Eq.~\eqref{eq:valence_neutron_wf}, the spin density for $p_{1/2}$ neutron orbital is given by
\begin{align}
	\psi_n^\dagger(\mathbf{x}_n) \bm{\sigma}_n \psi_n(\mathbf{x}_n) &= 
	\frac{R^2_{2p}(r_n)}{4\pi}
	\left[2 \left(\mathbf{n}_n \cdot\mathbf{n}_I\right) \mathbf{n}_n - \mathbf{n}_I\right].
\end{align}
The angular integral can be performed with the multipole expansion of the Coulomb potential
\begin{align}
	\frac{1}{\abs{\mathbf{x}_n - \mathbf{x}}} 
	&= \frac{\Theta(r_n-r)}{r_n} \sum_{l=0}^{\infty} \left(\frac{r}{r_n}\right)^l P_l (\cos \theta)
	+ \frac{\Theta(r-r_n)}{r} \sum_{l=0}^{\infty} \left(\frac{r_n}{r}\right)^l P_l (\cos \theta),
\end{align}
where $\cos \theta = \mathbf{x}_n \cdot \mathbf{x}/r r_n$,
and we obtain
\begin{align}
	e\mathbf{B}(\mathbf{x}) &=b_1(r) \mathbf{n}_I
	+ b_2(r) \left(3(\mathbf{n}\cdot\mathbf{n}_I)\mathbf{n} - \mathbf{n}_I\right),
\end{align}
where
\begin{align}
	b_1(r) &=\frac{\mu_n}{6\pi}  \left(2\int_r^\infty \frac{dr_n}{r_n}R_{2p}^2(r_n)-R_{2p}^2(r) \right),
	\quad
	b_2(r) = \frac{\mu_n}{12\pi}\left(R_{2p}^2(r) - \frac{1}{r^3} \int_0^r dr_n r_n^2R_{2p}^2(r_n)  \right),
\end{align}
thus reproducing the equations in the main text. As a cross check, one can show that these expressions satisfy Maxwell's equation $\nabla\cdot{\bf B}=0$.

In order to obtain $R_{2p}$,
we numerically solved the Schr\"odinger equation for the valence neutron moving in the Woods-Saxon potential. The parameters of the potential \cite{Dmitriev:2003sc} are tuned to reproduce single-particle energies and collective properties of heavy nuclei. We have checked that our numerical 
results are consistent with other single-particle calculations, of {\em e.g.} Schiff moment induced by the neutron EDM \cite{Ginges:2003qt}. Final numerical results for $S_{^{199}{\rm Hg}}$ are given in Eq. (\ref{numerical}).

%%%%%%%%%%%%%%%%%%%%%%%%%%%%%%%%%%%%%%%%%%%%%%%%
\section{Semi-leptonic $CP$-odd operator}
%%%%%%%%%%%%%%%%%%%%%%%%%%%%%%%%%%%%%%%%%%%%%%%%

Here we provide details on our evaluation of the semi-leptonic $CP$-odd operator $C_S$.
We again start from the $CP$-violating photon operator
\begin{align}
	\mathcal{L} &=
	\frac{e^4 C_{E^3B}}{8}(\tilde F_{\alpha\beta} F^{\alpha\beta})(F_{\gamma\delta}  F^{\gamma \delta}).
\end{align}
We contract two photons with the electron line as shown in Fig.~\ref{3loop}.
At the level of effective operators, this diagram is logarithmically divergent. However, since we have integrate out the muon,
the logarithmic divergence is tamed by the muon mass scale,
and hence we obtain
\begin{align}
	\mathcal{L} &= C_{E^3B} \times
	10\alpha^2 m_e\log\left(\frac{m_\mu}{m_e}\right) \abs{e\mathbf{E}}^2 \bar{e}i\gamma_5 e,
	\label{eq:CSequiv}
\end{align}
to the leading log accuracy,
where $\mathbf{E}$ is the nuclear electric field 
and we ignore $\mathbf{B}^2$ that is subdominant.
We use the same character $e$ for both gauge coupling and the electron spinor, but there should be no confusion.

It is well-known that the strength of atomic EDMs in heavy atoms is determined mostly by the mixing of $s_{1/2}$ and $p_{1/2}$ atomic orbitals. 
It is easy to see that both the ${\bf E}^2$-proportional  interaction~\eqref{eq:CSequiv} and the usual form of $C_S$-interaction~\eqref{eq:CS} induce
a mixing between the atomic $s_{1/2}$ and $p_{1/2}$ states.
Near/inside the nucleus where $\bar NN$ and ${\bf E}^2$ operators peak, the electron wave functions 
satisfy the Dirac equations and are  given by
\begin{align}
\label{eq:wf}
	\psi_{jlm} &=
	 e^{-iEt}
	\begin{pmatrix}
	 f_{jl}(r)\Omega_{jlm} \\
	 (-)^{j-l-1/2}g_{jl}(r)\Omega_{jl'm}
	 \end{pmatrix},
	\quad
	l' = 2j-l,
\end{align}
where $\Omega_{jlm}$ is the spherical harmonics spinor (see \emph{e.g.}~\cite{Berestetskii:1982qgu}).
Thus the atomic matrix element induced by~\eqref{eq:CS} is
\begin{align}
\label{eq:sp1}
	\int d^3x_N\, \rho_N(\mathbf{x}_N) \psi^\dagger_p(\mathbf{x}_N) \gamma^0 \gamma_5 \psi_s(\mathbf{x}_N)
	&= A\int dr_N\,r_N^2 \bar{\rho}_N \left(f_p g_s + f_s g_p\right),
\end{align}
where $\rho_N$ is the nucleon density distribution inside the nucleus, 
$A = 232$ is the atomic number of Th and we made the index $j = 1/2$ implicit for notational ease.

Atomic/molecular theory connects the small-$r$ asymptotic form of the wave functions (\ref{eq:wf}) with the full numerically determined atomic orbitals. $CP$-violation, on the other hand,  comes exclusively from the atomic short-distance matrix element (\ref{eq:sp1}). Therefore, in order to determine the atomic matrix element induced by~\eqref{eq:CSequiv} we need to replace (\ref{eq:sp1}) with 
\begin{align}
	\int d^3x_N\, \abs{e\mathbf{E}(\mathbf{x}_N)}^2 
	\psi^\dagger_p(\mathbf{x}_N) \gamma^0 \gamma_5 \psi_s(\mathbf{x}_N)
	&= \frac{24\pi Z^2 \alpha^2}{5R_N}
	\int dr_N\,r_N^2 \bar{\rho}_{E^2} \left(f_p g_s + f_s g_p\right).
	\label{eq:sp2}
\end{align}
Here the normalized distributions are taken as
\begin{align}
	\bar{\rho}_N(r_N) &\propto \frac{1}{1 + e^{(r_N-R_N)/a}},
	\quad
	\bar{\rho}_{E^2}(r_N) \propto \frac{r_N^2}{R_N^6}\Theta(R_N-r_N) + \frac{1}{r_N^4}\Theta(r_N-R_N),
	\quad
	\int dr_N r_N^2 \bar{\rho}_{N} = \int dr_N r_N^2 \bar{\rho}_{E^2} = 1.
\end{align}
Therefore the effective $C_S$-coupling induced by~\eqref{eq:CSequiv} is estimated as
\begin{align}
	\frac{G_F}{\sqrt{2}}C_S^{\mathrm{equiv}}  &= 
	C_{E^3B}\times \kappa \frac{48\pi Z^2 \alpha^4 m_e}{R_N A} 
	\log\left(\frac{m_\mu}{m_e}\right),
	\quad
	\kappa = \frac{\int dr_N\,r_N^2 \bar{\rho}_{E^2} \left(f_p g_s + f_s g_p\right)}
	{\int dr_N\,r_N^2 \bar{\rho}_N \left(f_p g_s + f_s g_p\right)}.
\end{align}
In order to evaluate the correction factor $\kappa$, that ultimately accounts for the difference of spatial distribution between $\bar NN$ and ${
\bf E}^2$ operators in the atomic matrix element, 
we solve the Dirac equation 
for the radial functions $f$ and $g$ numerically.
We take $R_N = r_0 A^{1/3}$ with $r_0 = 1.27\,\mathrm{fm}$
and $a = 0.742\,\mathrm{fm}$ following~\cite{Dmitriev:2003sc},
and obtain
\begin{align}
	\kappa \simeq 0.66.
\end{align}
This is used for our estimation of the upper limit on $d_\mu$ in the main text.

%%%%%%%%%%%%%%%%%%%%%%%%%%%%%%%%%%%%%%%%%%%%%%%%
\section{Comments on the accuracy of calculations}
%%%%%%%%%%%%%%%%%%%%%%%%%%%%%%%%%%%%%%%%%%%%%%%%

Since the calculation of $S(d_\mu)$ and $C_S(d_\mu)$ involve many steps, it is appropriate to comment on the expected accuracy of the results. The uncertainties can be subdivided into three categories, coming from particle physics, nuclear and atomic physics.

{\em Particle physics.} In calculating the muon loop leading to $({\bf BE}){\bf E}^2$ effective interaction, higher order terms in the electric field have been neglected. Such terms are additionally suppressed by powers of $(e{\bf E}^2)/m_\mu^4 \le (Z\alpha m_\mu^{-1}R_N)^4<10^{-3}$, and therefore this approximation holds really well. The loop integral also neglects the change of electric field on the scale of the muon Compton wavelength. This correction can be at maximum $\sim(R_N m\mu)^{-2}\sim 7\%$. Notice that this can be consistently improved by numerically calculating the muon loop in the realistic ${\bf E}(r)$ background. 

Photon loop calculation entering the calculation of $C_S$ has been performed to logarithmic accuracy, \emph{i.e.} 
$O(1)$ terms have been dropped relative to $\log(m_\mu^2/m_e^2)\sim O(10)$. This implies the accuracy of 10\%, which again can be improved upon numerical calculation of the two-loop (muon and photon) diagram. 

{\em Nuclear physics.} There are no nuclear uncertainties in the $C_S$ calculation, other than the exact modelling of the electric field distribution inside the nucleus. The charge distributions used in our calculations are ``anchored" by the measured values of the nuclear charge radii, but the exact shape can modeled by either constant-within-sphere, or Woods-Saxon form. This feeds into the calculation of the $\kappa$-factor, and we estimate that the possible variation does not exceed $\sim 10\%$. 

The calculation of the Schiff moment involves modelling of the magnetic field inside the nucleus. In our work it is done in the simple shell model that predicts the magnetic moment to be $\mu_{^{199}{\rm Hg}} = -\mu_n/3 = 0.637$, while in practice the measured result for this quantity is 0.509. The rest of the magnetism comes from the mixing of different nuclear orbital configurations, and neglecting it generates $\sim 20-25\%$ errors. It has to be emphasized that this uncertainty is much smaller than a very large, order of magnitude uncertainty in calculations of the Schiff moment induced by the $CP$-odd nuclear forces, where there is no valence contribution, and subtle effects in the core polarization widely vary as function of adopted nuclear models. 

{\em Atomic physics.} There is no change in atomic physics calculation (if the parameter $\kappa$ is treated as essentially a nuclear parameter). Therefore, same atomic calculations of molecular/atomic orbitals apply, and modern calculations are performed with estimated errors not exceeding 10\%. 

Combining different sources of errors, we conclude that the calculation of $C_S(d_\mu)$ and the resulting bounds on $d_\mu$ carry a theoretical error of $\sim 15-20\%$ which can be brought down to 10\% level with more accurate modelling of the nuclear electric field distribution and full calculation of the two-loop diagram. Calculation of $S(d_\mu)$ carries a $\sim 30\%$ uncertainty, mostly due to our reliance on the simple shell model, but can be improved with a more sophisticated nuclear input.

%%%%%%%%%%%%%%%%%%%%%%%%%%%%%%%%%%%%%%%%%%%%%%%%
\end{document}